# Quantum Hall *p-n* Junction Dartboards Using Graphene Annuli

C.-I Liu,[1] D. K. Patel,[1,2] M. Marzano,[1,3,4] M. Kruskopf,[1,5] H. M. Hill,[1] and A. F. Rigosi[1,a]

[1]*Physical Measurement Laboratory, National Institute of Standards and Technology (NIST), Gaithersburg, Maryland, 20899-8171, USA*

[2]*Department of Physics, National Taiwan University, Taipei, 10617, Taiwan*

[3]*Department of Electronics and Telecommunications, Politecnico di Torino, Torino, 10129, Italy*

[4]*Istituto Nazionale di Ricerca Metrologica, Torino, 10135, Italy*

[5]*Joint Quantum Institute, University of Maryland, College Park, MD 20742, USA*

The use of multiple current terminals on millimeter-scale graphene *p-n* junction devices fabricated with Corbino geometries, or quantum Hall resistance dartboards, have enabled the measurement of several fractional multiples of the quantized Hall resistance at the $\nu = 2$ plateau ($R_H \approx 12906$ $\Omega$). Experimentally obtained values agreed with corresponding numerical simulations performed with the LTspice circuit simulator. More complicated designs of the quantum Hall resistance dartboard were simulated to establish the potential parameter space within which these Corbino-type devices could output resistance. Most importantly, these measurements support simpler processes of ultraviolet lithography as a more efficient means of scaling up graphene-based device sizes while maintaining sufficiently narrow junctions.

---

[a] Author to whom correspondence should be addressed.  Mail: Albert Rigosi, MS 8171, 100 Bureau Drive, NIST, Gaithersburg, MD 20899.

Devices fabricated from graphene have been the subject of heavy research since the honeycomb-lattice structure was discovered [1-4]. In the regime of the quantum Hall effect (QHE), graphene exhibits quantized resistance values at $\frac{1}{(4n+2)}R_\text{K}$, where $R_\text{K} = \frac{h}{e^2}$ (also known as the von Klitzing constant), *n* equals an integer, *e* is the elementary charge, and *h* is the Planck constant. Traditional Hall devices that contain *p-n* junctions (*pn*Js) also exhibit various multiples and fractions of the von Klitzing constant when studying its transport properties in the QHE [5-14], as do devices that take on a Corbino geometry [15-17]. The fabrication size of general graphene devices is essential for large scale production of thin films [18-19]. Moreover, the demonstration of functional *pn*Js on this large of a scale may have ramifications for photodetection [20-23] and electron optics [24-27]. Some specific applications of these Corbino devices, much like those that will be presented, include the construction of a curved two-dimensional Dirac fermion microscope [28], custom programmable quantized resistors [29], and mesoscopic valley filters [30].

Since graphene has been at the core of many prospects, commercial and otherwise, the pursuit of accommodating large-scale *pn*J devices is still a relevant endeavor. Here, large-scale is meant to be of centimeter order or larger, and to date, creating such *pn*J devices remains tricky if fabrication is attempted by using electrostatic gates, which may easily permit leakage currents on this length scale. The second factor, regarding access to quantized resistance values, has been heavily explored though analyses on Landauer-Büttiker edge state equilibration [5-7, 31-34]. Though this approach provides one way of accessing different quantized values, it does so with the condition that various regions of a *pn*J device are held at Landau levels that need not be equal, and this is typically accomplished by using an electrostatic gate to tune the Fermi level. Access to different quantized values of resistance can be greatly simplified if alternative approaches are explored. As shown in recent studies [35-36], one approach involves the use of multiple current terminals, which largely opens the quantized resistance parameter space achievable with traditional (linear) Hall *pn*J devices, and possibly other device geometries. The key advantage in using the Corbino geometry over the traditional Hall bar is that many more quantized values can potentially be obtained due to the periodic boundary conditions imposed on the electron flow.

In this work, centimeter-scale fabrication of epitaxial graphene (EG) *pn*J devices having a Corbino geometry is demonstrated, along with the added benefit of obtaining sufficiently sharp junctions from using standard ultraviolet photolithography (UVP) and ZEP520A [35]. These EG Corbino *pn*J devices can take on quantized Hall resistance values, in part, because of the junctions' provision to allow edge-state electrons to flow from the inner to the outer edges of the annulus or vice versa. Furthermore, devices were verified with quantum Hall transport measurements and LTspice current simulations [37] for cases where multiple current terminals were used. Overall, these experiments serve to support and validate three main points: the scalability of *pn*J devices, the versatility of *pn*J circuits using multiple current terminals, and the flexibility provided by large-scale junctions in transforming devices with Corbino geometries into those that allow edge-state current flow between the two edges, or quantum Hall resistance dartboards.

The simulations for the Corbino *pn*J devices were performed with the electronic circuit simulator LTspice, with quantum Hall elements that were used as described in relevant works [38, 39]. The simulated circuit utilizes both *p*-type and *n*-type quantum Hall elements [38], and both are respectively designated as either having ideal counterclockwise (CCW) or clockwise (CW) edge-state current flow. To summarize Section 2 of Ref. [38], quantum Hall elements are represented as objects that are compatible with circuit theory. Such treatment can be derived from Landauer-Büttiker formalism [40]. EG was grown on centimeter-sized SiC chips that were diced from 4*H*-SiC(0001) wafers (CREE) [41]. All cleaning and treatments performed before the growth, namely the processing with AZ5214E to utilize polymer-assisted sublimation, are well documented [42]. The growth was performed in an argon environment at 1900 °C using a graphite-lined resistive-element furnace (Materials Research Furnaces Inc.) [41] with heating and cooling rates of about 1.5 °C/s.

Once the EG is fully grown, samples were assessed with confocal laser scanning and optical microscopy. After inspection of monolayer growth, devices were fabricated, using Pd and Au as protective layers against organic contamination [43]. To improve contact resistances for cryogenic temperatures, electrical contacts were composed of NbTiN, a known superconductor with a $T_c$ of about 10 K at 9 T [44-45]. Upon completing the Corbino *pn*J devices, a functionalization treatment using $Cr(CO)_3$ was performed for two primary reasons: (1) to reduce the electron density to approximately $10^{10}$ cm$^{-2}$ and (2) to provide uniformity along the centimeter-scale devices [46-48]. The functionalization enables a relative ease of control of the electron density by annealing [49]. After this step, the wafer containing many devices is diced into smaller chips for further processing, with each device having a lateral size of millimeter scale. For the formation of the *pn*Js, S1813 photoresist was deposited as a spacer layer for intended *n*-type regions. Standard UVP was used to selectively etch regions intended for *p*-type adjustment, subsequently followed by deposition of mixed poly-methyl methacrylate and methyl methacrylate (PMMA/MMA) as the mediation layer for ZEP520A, deposited as the photoactive layer [35, 49].

When fabrication was complete, devices were exposed to UV radiation of wavelength 254 nm, and with similar timescales used as Ref. [35], *p*-type doping was achieved in regions not protected by the S1813 spacer layer. Because the *n*-type regions were unaffected by the UV exposure, they retained their initial electron density after functionalization, of the order $10^{10}$ cm$^{-2}$. Only after annealing the devices for approximately 20 min to 30 min at a temperature of 350 K would the *n*- type regions exhibit high enough electron density to form the quantized plateau at ν = 2. Example measurements are shown in Fig. 1.

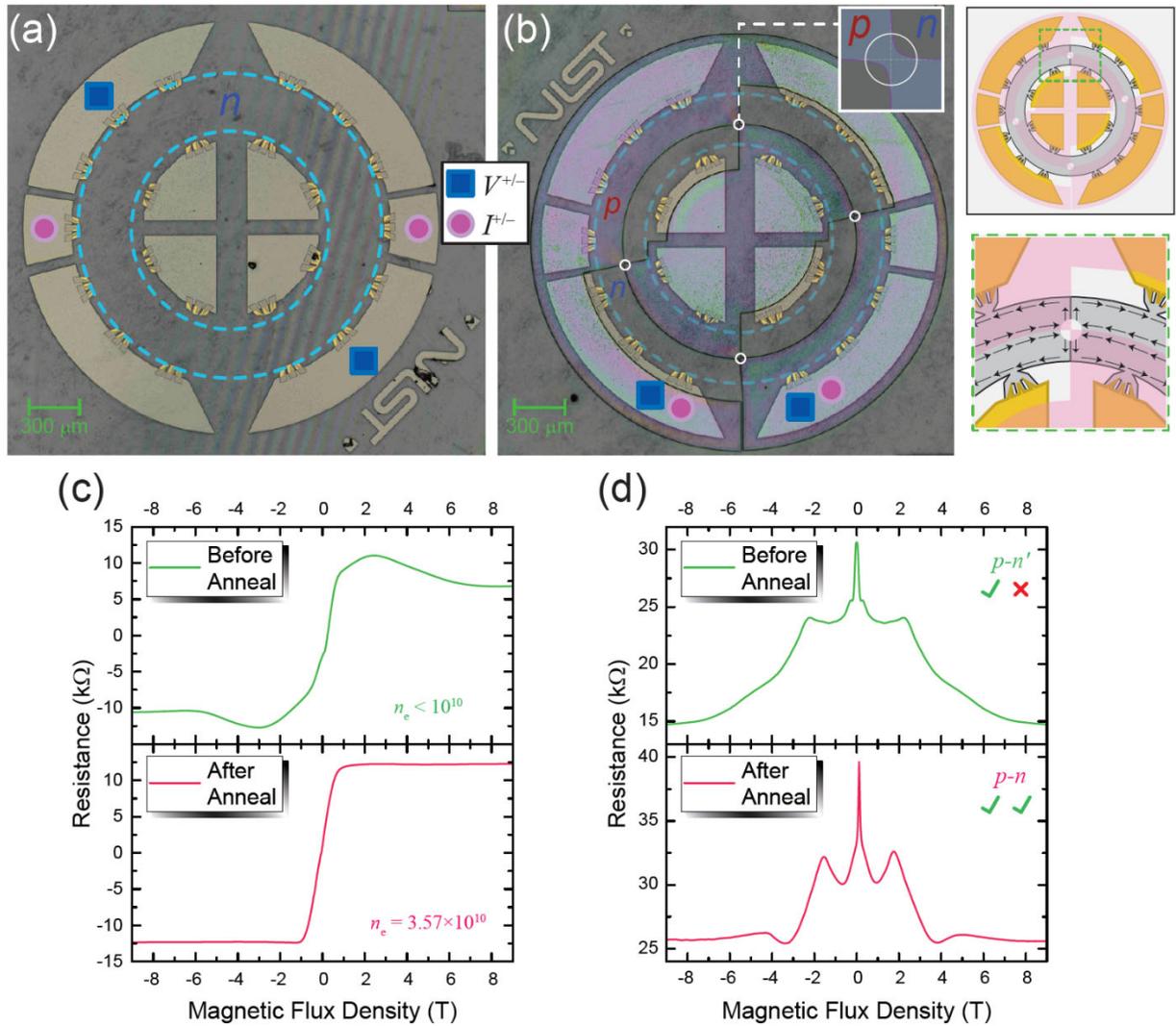

FIG. 1. (a) A control device with overlaid markings in a dashed light blue color to indicate the approximate bounds of the graphene annulus. The device was *n*-type and was exposed to the same UV conditions as the experimental devices. Pink dots and blue squares indicate the terminals for current injection and voltage measurements, respectively. (b) An example experimental device prior to wire bonding. Regions that appear with a lighter, bluish shade have had the S1813 layer removed, enabling the subsequently deposited PMMA/MMA and ZEP-520A to provide electron acceptors close enough to the graphene surface, enabling *p*-type doping. White circles indicate approximate areas where graphene was removed to enhance region separation, as illustrated in the top-right inset. The device schematic is also provided to the right of the image, with a full view on the top (graphene in gray, intended *p*-type region with shaded pink) and a magnified view below. The latter is shown to provide clarity on the direction of edge currents within the device. (c) The Hall resistance was determined for the control device in the cases before and after annealing. After annealing, the *n*-type region exhibited quantization. (d) A different test was necessary for the experimental device, whereby the annealing treatment would activate a working *p-n* junction between the two blue squares in (b).

Control devices accompanied the experimental Corbino *pn*J devices as a means to monitor the general behavior of *n*-type regions using traditional Hall measurements. An example control device is shown in Fig. 1 (a). The image is overlaid with dashed light blue circles indicating the approximate boundaries of the graphene annulus. The millimeter-scale, *n*-type devices were subjected to the same UV exposure as their experimental counterparts. Pink dots and blue squares show where current and voltage terminals were for characterization measurements, respectively. In Fig. 1 (b), one of the experimental devices is shown prior to wire bonding. The regions that appear with a lighter, more bluish shade are the regions that have no S1813 spacer layer and were intended to be *p*-type. The lack of the spacer layer enabled the photoactive ZEP-520A to attract electrons from the graphene surface [35, 49]. During the initial fabrication steps, when the graphene annulus was formed, small holes with diameters of about 1 μm were also etched in four spots, as represented by white circles in the inset of Fig. 1 (b). These small spots were anticipated to encircle the intersection of four adjacent regions. In the event that the lithographic process was insufficient to provide high enough resolution to create a perfect set of four corners, then, as shown in the same inset, two of the regions with the same polarity would be connected, thus altering the intended geometry of the devices. By removing graphene from this region, the intended geometry was more easily realized.

To verify the functionality of the devices, various tests were performed. First, the Hall resistances of the control devices were determined in cases before and after annealing, as seen in Fig. 1 (c). Since each control device had an experimental counterpart, the same conditions were applied to the latter, and the corresponding test data are shown in Fig. 1 (d). With the assistance of LTspice simulations, two-terminal measurements were compared on the experimental devices. Prior to annealing, when the *p*-type regions could exhibit quantization at 9 T but the *n*-type regions could not, the two-terminal measurement (shown in the top panel) did not output the expected value of $R_K$. However, after annealing, the *n*-type region exhibited quantization at the ν = 2 plateau ($\frac{R_K}{2} = R_H \approx 12906$ Ω) for the control devices (bottom panel of Fig. 1 (c)). By applying the same annealing conditions to the experimental devices, the two-terminal measurement illustrated in Fig. 1 (b) began outputting the expected quantized resistance of $R_K = 2 * R_H$ (where $R_H$ is being used as the relative unit of quantized resistance). This correct output was an indication that the annealing treatment had activated a working *pn*J between the two blue squares in Fig. 1 (b). Two-terminal measurements for the other neighboring pairs yielded the same result.

With functioning millimeter-sized Corbino *pn*J devices, the next step was to show that these devices could support the multiple-terminal approach for obtaining different quantized values of resistance. We define $R_{AB} = q * R_H$, where *q* is the coefficient of effective resistance (CER). This number can be simulated by LTspice with the schematic shown in Fig. 2 (a). Provided that the voltage was always measured at points A and B (that is, prior to splitting the source or drain into arbitrary branches), the CER could always be well-defined. In this example arrangement of having a dual-split source, the CER was simulated to be $q = \frac{3}{2}$.

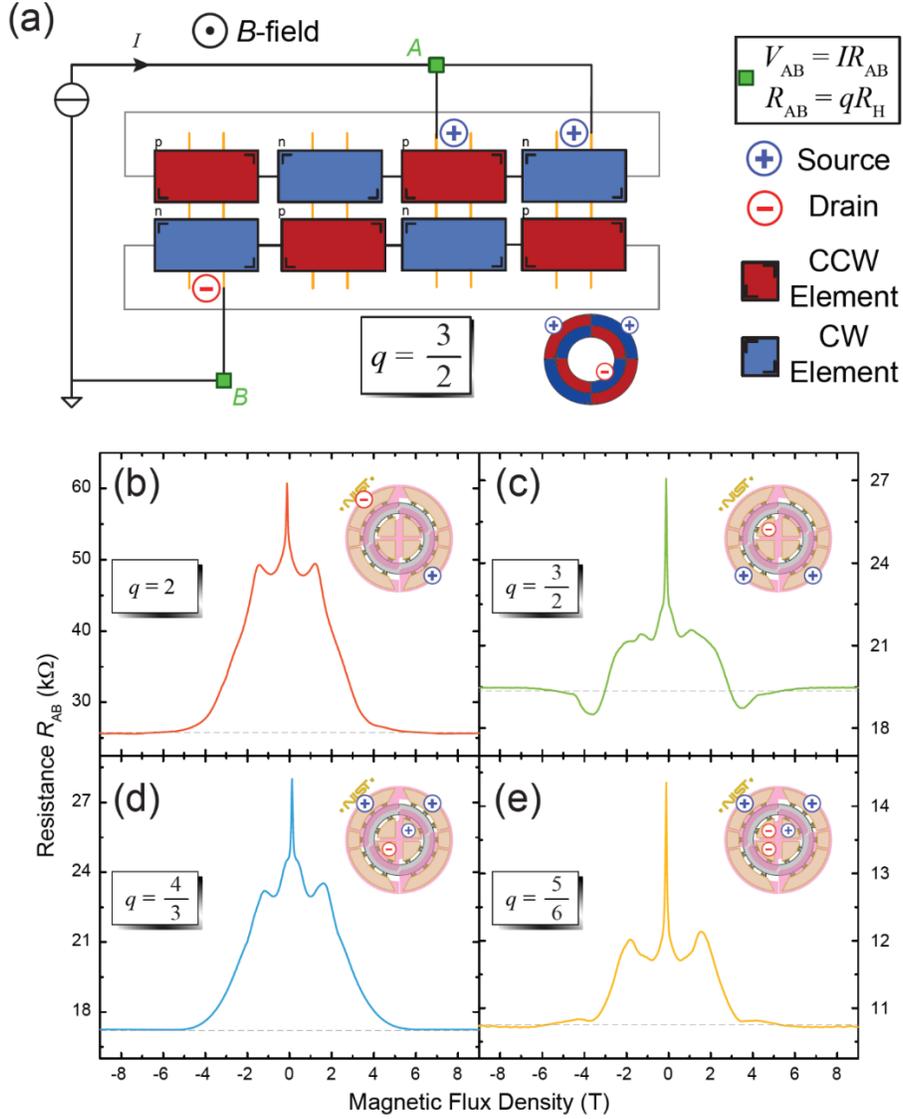

FIG. 2. (a) An example schematic of the LTspice simulation. The effective voltage, measured between A and B, gives the effective resistance of the device and is expressed as a multiplicative factor ($q$) times the Hall resistance $R_H$. (b) – (e) The resistance data from experimental devices were obtained by using multiple source and drain terminals (plus and minus, respectively). Depending on the arrangement of the terminals used for current injection (top right illustration within each panel), different values of $q$ were realized. The dashed gray lines in each panel correspond to the simulated $q$ value obtained from LTspice.

Simulations were then performed for several cases that yielded distinct CERs. In Fig. 2 (b), with the number of terminals being defined as $N$, a two-terminal measurement was performed again for a different pair, agreeing with the simulated value $q = 2$, which is shown as a dotted gray line in all device measurements. In Fig. 2 (c), an $N = 3$ arrangement was made with two sources and a single drain, yielding $q = \frac{3}{2}$ like Fig. 2 (a). The arrangements are illustrated in the top right corner of every measurement panel along with the corresponding sources and drains, represented by blue pluses and red minuses, respectively. For the $N = 4$ case shown in Fig. 2 (d) (three sources, one drain), the simulation yielded a CER of $q = \frac{4}{3}$. The corresponding experimental data agree with this predicted value. In the final case of Fig. 2 (e), with $N = 5$ (three sources, two drains), the CER was simulated to be $q = \frac{5}{6}$ and also agreed with its experimental counterpart. These values can be numerically calculated with Landauer-Büttiker formalism by arranging a matrix equation with pseudo-contacts at each distinct region of the schematic in Fig. 2 (a) [5, 6, 40]. As described in more detail in the aforementioned references, the currents and potentials can be determined at each of the 20 sides, requiring a lengthy transmission matrix that can be partially simplified with conditions imposed by Kirchoff's laws and the charge conservation law [5, 6, 40]. An example of calculated voltages for Fig. 2 (c) is provided in the Supplementary Material.

The data acquired on these devices support the achievable goal that, in addition to being able to scale up *pn*J devices, one may also use multiple terminals to obtain different quantized resistances, as well as utilize the Corbino geometry in the event that linear *pn*J devices are unable to provide particular fractions that may be more easily accessible in this parameter space. This advantage of accessing more varied quantized values may justify the use of the Corbino geometry over the linear one. The varied access arises due to the imposed periodic boundary conditions on the electronic potential and thus electron flow of the device.

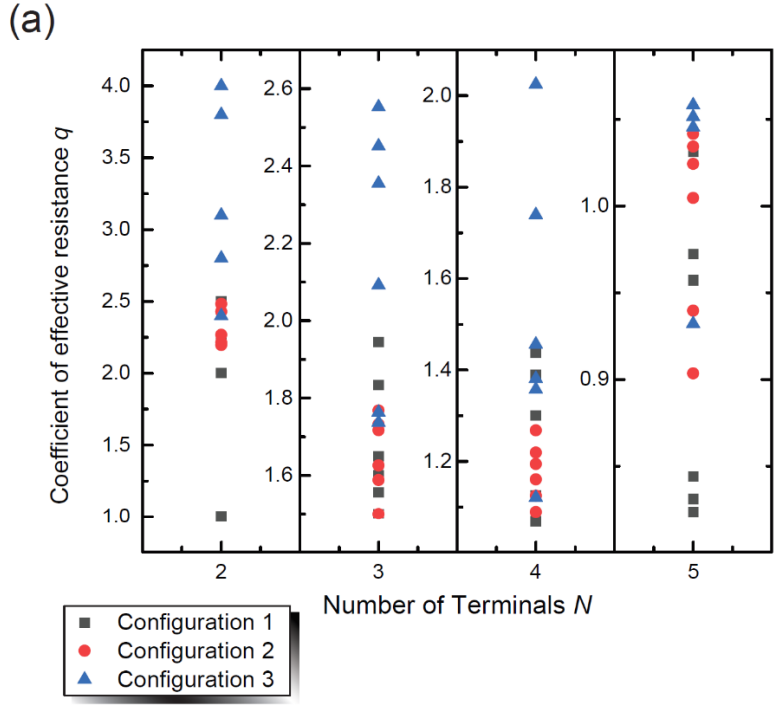

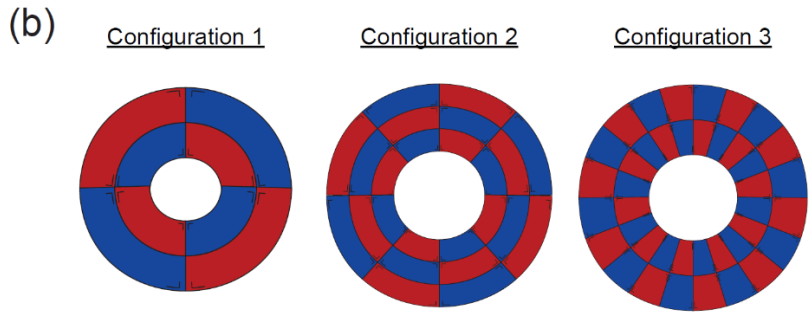

FIG. 3. (a) The data from LTspice simulations are shown for three configurations of proposed Corbino *pn*J devices. Configurations 1, 2, and 3 are represented by black squares, red circles, and blue triangles, respectively, for a subset of all possible cases using a varying number of terminals as sources or drains. For each situation that uses a certain number of terminals, the vertical scale gives some idea of how many quantized resistances become accessible without the need for electrostatic gating. All fractions represented graphically are listed in the Supplementary Material. (b) An illustration of each of the configurations is provided for clarity. The first configuration has two layers and four junction members, whereas the second contains three layers and eight junction members. The third configuration, as proposed, would contain two layers and twenty junction members, reflecting an embodiment of a true dartboard.

With the device fabrication able to result in functioning devices, one may customize device geometries to obtain particular fractions of $R_\text{H}$. In Fig. 3 (a), simulation data are shown for three distinct quantum Hall resistance dartboards and represent many of the achievable CERs as a function of how many terminals are used. The total range of CERs bears some dependence on the number of junction members along the Corbino *pn*J device perimeter, whereas the number of terminals used for a single measurement seems to enable fine-tuning the obtainable value within a smaller neighborhood of possible quantized resistances. Configurations 1, 2, and 3 are represented by black squares, red circles, and blue triangles, respectively. Only a randomly-selected subset of all possible values for each configuration has been graphically represented for the sake of highlighting the range of accessible values. Furthermore, simulations that utilized two, three, four, and five terminals have been assigned four different vertical axes for clarity. All fractions that are represented graphically can be found in the Supplementary Material. Fig. 3 (b) shows an illustration of each of the configurations. The first configuration has two layers and four junction members, the second contains three layers and eight junction members, and the third contains two layers and twenty junction members. The latter configuration most closely resembles an embodiment of a true dartboard. As *pn*J device applications continue to evolve, it becomes of more beneficial and relevant knowledge that Corbino geometries can be fabricated for quantum Hall measurements while also providing an alternative set of accessible quantized resistances to its linear, more traditional Hall bar counterpart.

In conclusion, this work pursued the validation and achievement of three related goals: the fabrication of Corbino *pn*J devices on a centimeter-scale, the ability to use multiple terminals for accessing different quantized resistance values, and the ability to transform a device with Corbino geometry into one that can generate resistance plateaus by enabling edge-state current flow between both edges. The experimental data for the quantized values matched the LTspice simulations, thus validating the use of the simulation software as a means to propose more complex devices for realizing an abundance of effective quantized resistances in *pn*J circuits.

## SUPPLEMENTARY MATERIAL

See supplementary material for a table of fractional values graphically represented in Fig. 3 (a) and for an example calculation of voltages in Fig. 2. The data that support the findings of this study are available from the corresponding author upon reasonable request.

## ACKNOWLEDGMENTS

The work of DKP at NIST was made possible by C-T Liang of National Taiwan University. The work of MM at NIST was made possible by M Ortolano of Politecnico di Torino and L Callegaro of Istituto Nazionale di Ricerca Metrologica, and the authors thank them for this endeavor. The authors would like to express thanks to L Chao and A Levy for their assistance in the NIST internal review process.

# Supplementary Online Material: Quantum Hall resistance dartboards using graphene *p-n* junction devices with Corbino geometries

C.-I Liu,[1] D. K. Patel,[1,2] M. Marzano,[1,3,4] M. Kruskopf,[1,5] H. M. Hill,[1] and A. F. Rigosi,[1,a)]

[1]*Physical Measurement Laboratory, National Institute of Standards and Technology (NIST), Gaithersburg, Maryland, 20899-8171, USA*

[2]*Department of Physics, National Taiwan University, Taipei, 10617, Taiwan*

[3]*Department of Electronics and Telecommunications, Politecnico di Torino, Torino, 10129, Italy*

[4]*Istituto Nazionale di Ricerca Metrologica, Torino, 10135, Italy*

[5]*Joint Quantum Institute, University of Maryland, College Park, MD 20742, USA*

Table of Contents:

1. Table of Fractions

2. Calculating Voltage Elements (Example)

## 1. Table of Fractions

TABLE S1. This table provides exact fractional values of data represented in Fig. 3 (a) of the main text.

|       | Configuration 1 | Configuration 2 | Configuration 3 |
|-------|-----------------|-----------------|-----------------|
| $N = 2$ | 2     | 123/56    | 19/5   |
|       | 9/4   | 127/56    | 12/5   |
|       | 5/2   | 139/56    | 4      |
|       | 1     | 17/7      | 14/5   |
|       | --    | 31/14     | 31/10  |
| $N = 3$ | 3/2   | 2981/1736 | 33/19  |
|       | 14/9  | 3/2       | 67/38  |
|       | 8/5   | 97/56     | 159/76 |
|       | 11/6  | 27/17     | 179/76 |
|       | 35/18 | 387/238   | 97/38  |
|       | 33/20 | 99/56     | 76/31  |
| $N = 4$ | 16/15 | 71/56     | 164/81 |
|       | 9/8   | 1707/1400 | 327/188 |
|       | 46/43 | 65/56     | 150/103 |
|       | 13/10 | 9/8       | 337/244 |
|       | 25/18 | 61/56     | 167/123 |
|       | 23/16 | 1137/952  | 139/124 |
| $N = 5$ | 33/32 | 25/24     | 563/532 |
|       | 35/36 | 3613/3596 | 551/524 |
|       | 67/70 | 6601/6444 | 409/356 |
|       | 14/17 | 6601/7025 | 961/1031 |

| | | |
|---|---|---|
| 65/77 | 1476/1427 | 23/22 |
| 59/71 | 234/259 | 1101/836 |

## 2. Calculating Voltage Elements (Example)

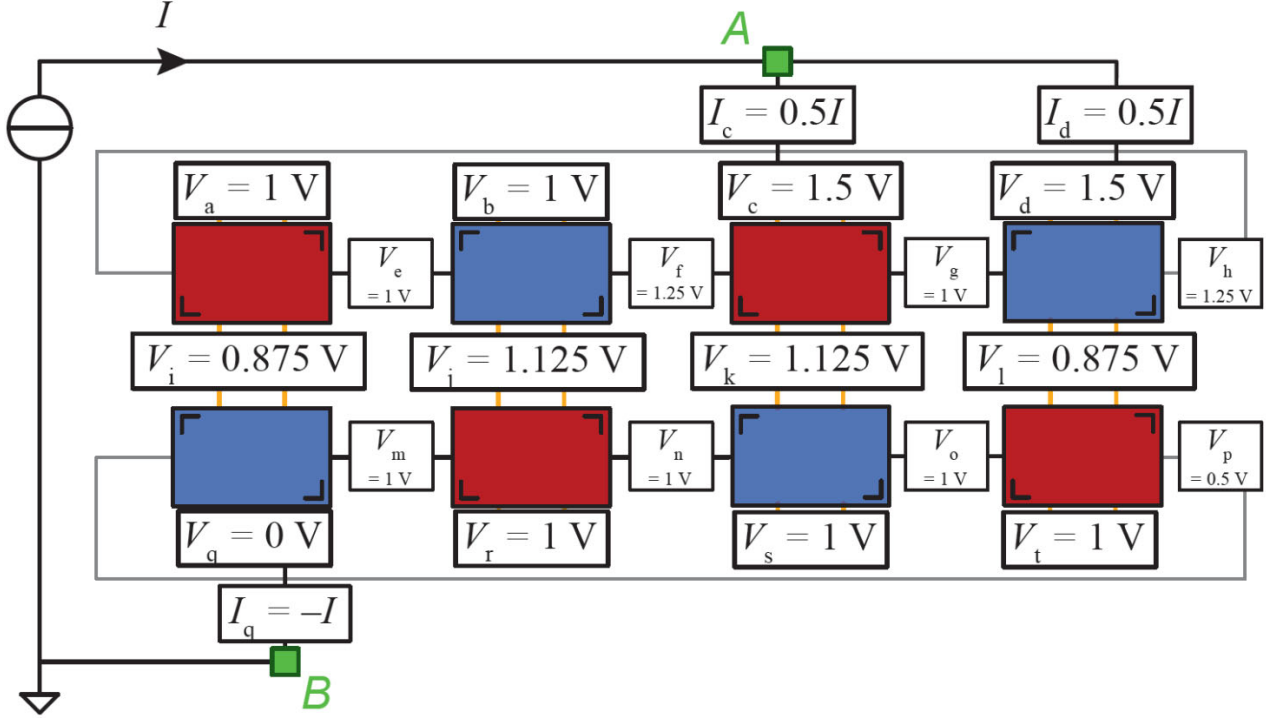

FIG. S1. The voltage elements and current branches are determined numerically for the case of Fig. 2 (c). For further analysis details on using transmission matrices, please consult main text Refs. [5, 6, 40]. The matrix analysis can begin with 20 elements if contacts are assumed at each interface. Some simplifications can be made by conservation of current. The conductance quantum at the $\nu = 2$ plateau ($\frac{h}{2e^2}$) is set to one for this figure.

For the corresponding matrix analysis, each node conserves current, and for many elements, that net value is zero since current is only allowed to exit the device through one contact at the lower left corner of Fig. S1. Taking the matrix equation, **I = GV**, we solve for V to get:

$$
\begin{bmatrix} 0 \\ 0 \\ 1/2 \\ 1/2 \\ 0 \\ 0 \\ 0 \\ 0 \\ 0 \\ 0 \\ 0 \\ 0 \\ 0 \\ 0 \\ 0 \\ 0 \\ -1 \\ 0 \\ 0 \\ 0 \end{bmatrix}
=
\begin{bmatrix}
1 & 1 & 0 & 0 & -2 & 0 & 0 & 0 & 0 & 0 & 0 & 0 & 0 & 0 & 0 & 0 & 0 & 0 & 0 & 0 \\
0 & 1 & 1 & 0 & 0 & -2 & 0 & 0 & 0 & 0 & 0 & 0 & 0 & 0 & 0 & 0 & 0 & 0 & 0 & 0 \\
0 & 0 & 1 & 1 & 0 & 0 & -2 & 0 & 0 & 0 & 0 & 0 & 0 & 0 & 0 & 0 & 0 & 0 & 0 & 0 \\
0 & 0 & 0 & 1 & 1 & 0 & 0 & -2 & 0 & 0 & 0 & 0 & 0 & 0 & 0 & 0 & 0 & 0 & 0 & 0 \\
0 & 0 & 0 & 0 & 1 & 0 & 0 & 0 & -1 & -1 & 0 & 0 & 1 & 0 & 0 & 0 & 0 & 0 & 0 & 0 \\
0 & 0 & 0 & 0 & 0 & 1 & 0 & 0 & 0 & -1 & -1 & 0 & 0 & 1 & 0 & 0 & 0 & 0 & 0 & 0 \\
0 & 0 & 0 & 0 & 0 & 0 & 1 & 0 & 0 & 0 & -1 & -1 & 0 & 0 & 1 & 0 & 0 & 0 & 0 & 0 \\
0 & 0 & 0 & 0 & 0 & 0 & 0 & 1 & 0 & 0 & 0 & -1 & -1 & 0 & 0 & 1 & 0 & 0 & 0 & 0 \\
0 & 0 & 0 & 0 & -1 & 0 & 0 & 0 & 1 & 1 & 0 & 0 & -1 & 0 & 0 & 0 & 0 & 0 & 0 & 0 \\
0 & 0 & 0 & 0 & 0 & -1 & 0 & 0 & 0 & 1 & 1 & 0 & 0 & -1 & 0 & 0 & 0 & 0 & 0 & 0 \\
0 & 0 & 0 & 0 & 0 & 0 & -1 & 0 & 0 & 0 & 1 & 1 & 0 & 0 & -1 & 0 & 0 & 0 & 0 & 0 \\
0 & 0 & 0 & 0 & 0 & 0 & 0 & -1 & 1 & 0 & 0 & 1 & 0 & 0 & 0 & -1 & 0 & 0 & 0 & 0 \\
0 & 0 & 0 & 0 & 1 & 0 & 0 & 0 & -1 & -1 & 0 & 0 & 1 & 0 & 0 & 0 & 0 & 0 & 0 & 0 \\
0 & 0 & 0 & 0 & 0 & 1 & 0 & 0 & 0 & -1 & -1 & 0 & 0 & 1 & 0 & 0 & 0 & 0 & 0 & 0 \\
0 & 0 & 0 & 0 & 0 & 0 & 1 & 0 & 0 & 0 & -1 & -1 & 0 & 0 & 1 & 0 & 0 & 0 & 0 & 0 \\
0 & 0 & 0 & 0 & 0 & 0 & 0 & 1 & -1 & 0 & 0 & -1 & 0 & 0 & 0 & 1 & 0 & 0 & 0 & 0 \\
0 & 0 & 0 & 0 & 0 & 0 & 0 & 0 & 0 & 0 & 0 & -2 & 0 & 0 & 0 & 1 & 1 & 0 & 0 & 0 \\
0 & 0 & 0 & 0 & 0 & 0 & 0 & 0 & 0 & 0 & 0 & 0 & -2 & 0 & 0 & 0 & 1 & 1 & 0 & 0 \\
0 & 0 & 0 & 0 & 0 & 0 & 0 & 0 & 0 & 0 & 0 & 0 & 0 & -2 & 0 & 0 & 0 & 1 & 1 & 0 \\
0 & 0 & 0 & 0 & 0 & 0 & 0 & 0 & 0 & 0 & 0 & 0 & 0 & 0 & -2 & 1 & 0 & 0 & 0 & 1
\end{bmatrix}
\begin{bmatrix} V_a \\ V_b \\ V_c \\ V_d \\ V_e \\ V_f \\ V_g \\ V_h \\ V_i \\ V_j \\ V_k \\ V_l \\ V_m \\ V_n \\ V_o \\ V_p \\ V_q \\ V_r \\ V_s \\ V_t \end{bmatrix}
$$

(S1)

$$
V = \begin{bmatrix} 1 \\ 1 \\ 1.5 \\ 1.5 \\ 1 \\ 1.25 \\ 1 \\ 1.25 \\ 0.875 \\ 1.125 \\ 1.125 \\ 0.875 \\ 1 \\ 1 \\ 1 \\ 0.5 \\ 0 \\ 1 \\ 1 \\ 1 \end{bmatrix}
$$

(S2)